\documentclass[twocolumn,preprintnumbers,amsmath,amssymb,showpacs]{revtex4}

\usepackage{graphicx}
\usepackage{dcolumn}
\usepackage{bm}

\begin{document}

\title{Enumeration of distinct mechanically stable disk packings in 
small systems}

\author{Guo-Jie Gao$^1$}
\author{Jerzy Blawzdziewicz$^1$} 
\author{Corey S. O'Hern$^{1,2}$}
\affiliation{$^1$~Department of Mechanical Engineering, Yale University, 
New Haven, CT 06520-8284.\\
$^2$~Department of Physics, Yale University, New Haven, CT 06520-8120.\\
}

\date{\today}

\begin{abstract}
We create mechanically stable (MS) packings of bidisperse disks using
an algorithm in which we successively grow or shrink soft repulsive
disks followed by energy minimization until the overlaps are
vanishingly small.  We focus on small systems because this enables us
to enumerate nearly all distinct MS packings.  We measure the
probability to obtain a MS packing at packing fraction $\phi$ and find
several notable results.  First, the probability is highly nonuniform.
When averaged over narrow packing fraction intervals, the most
probable MS packing occurs at the highest $\phi$ and the probability
decays exponentially with decreasing $\phi$.  Even more striking,
within each packing-fraction interval, the probability can vary by
many orders of magnitude.  By using two different packing-generation
protocols, we show that these results are robust and the packing
frequencies do not change qualitatively with different protocols.
\end{abstract}

\maketitle

\section{Introduction}
\label{introduction}

Inherent structures or potential energy minima are important for
determining the mechanical and dynamical properties of supercooled
liquids and glasses \cite{stillinger}.  Recently, a host of
computational studies have attempted to relate thermodynamic
quantities in supercooled liquids to the number of inherent structures
and vibrational motions about them \cite{shell,nave,shell2}.  It is
often assumed in these calculations that all inherent structures at a
given energy are equally probable.  Similarly, statistical
descriptions of granular media assume that all stable particle
packings at a particular volume are equally likely
\cite{edwards,makse2}.  Thus, it is important to examine under what
conditions and to what extent that inherent structures in glassy
systems and jammed packings of granular materials are equally
probable.  In this short article, we begin to address this question by
enumerating nearly all mechanically stable (MS) packings in small 2d
bidisperse systems.
 
An innovative feature of this work and our other recent study
\cite{xu} is that we focus on systems containing small numbers of
disks.  We note that related studies of small hard disk systems have
been carried out previously, but these have not investigated the
MS packing probabilities \cite{bs,bsv}.  We confine our studies to
small 2d systems for two key reasons.  First, we are able to generate
nearly all of the mechanically stable disk packings in these systems.
The number of MS disk packings grows exponentially with the number of
particles $N$, but is finite for any finite $N$.  In small
systems, we are able to accurately determine the probability with
which each MS packing occurs.  In contrast, in large systems, only the
most frequent MS packings are found.  Second, we believe that
understanding small jammed systems is crucial to developing a
theoretical explanation for slow stress and structural relaxation
in large glassy and amorphous systems \cite{adam}.  Our view of
the relationship between jamming in small systems and glassy behavior
in large systems will be more fully developed in a future publication
\cite{gg}.

\begin{figure}
\centerline{\scalebox{0.5}{\includegraphics{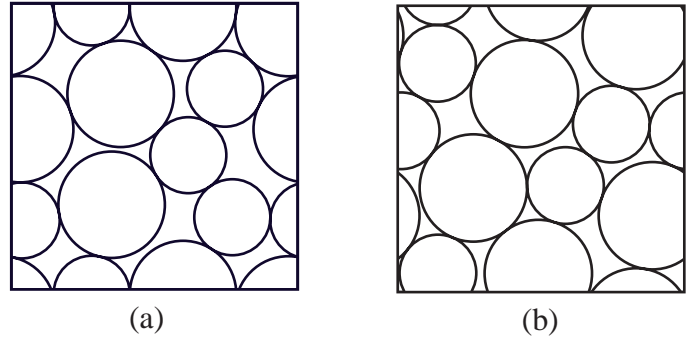}}}%
\vspace{-0.1in}
\caption{\label{fig:compare} Snapshot of (a) a MS packing at $\phi
\approx 0.82$ and (b) another MS packing $\phi \approx 0.83$ that is
$10^6$ times more frequent.}
\vspace{-0.1in}
\end{figure}

We discuss several important results below. First, we find that the
frequency with which mechanically stable packings occur is not
uniform, instead it can vary by many orders of magnitude over the
range of packing fractions where MS packings exist.  This large
variation in frequency occurs even though we do not
target specific packings.  Moreover, MS packings can
have frequencies that differ by many orders of magnitude even over
narrow packing-fraction intervals and there are no striking
structural differences between rare and frequent MS packings at
similar $\phi$ as shown in Fig.~\ref{fig:compare}.  Thus, it is not immediately
obvious what variables set the frequency of MS packings.  Also, we
show below that even when we significantly alter the protocol used to
generate the MS packings, the most frequent packings remain frequent
and the rare ones remain rare.  Thus, we suggest that although it is
clear that the particular protocol chosen to generate the MS packings
plays some role in determining the frequency distribution of MS
packings\cite{torquato2}, prominent geometrical features of the
configuration space can also strongly influence the frequency
distribution.

\vspace{-0.1in}

\section{Simulation Methods}
\label{methods}

\subsection{Generation of Mechanically Stable Packings}
\label{generation}

We study mechanically stable packings of $N=10$ disks that interact via
the finite-range, pairwise additive, purely repulsive spring potential
\begin{equation}
\label{spring potential}
V(r_{ij})
=\frac{\epsilon}{2}(1-r_{ij}/\sigma_{ij})^2\Theta(\sigma_{ij}/r_{ij}-1).
\end{equation}
Here $\epsilon$ is the characteristic energy scale, $r_{ij}$ is the
separation between particles $i$ and $j$, $\sigma_{ij}=\left(\sigma_i
+ \sigma_j \right) /2$ is their average diameter, and $\Theta(x)$ is
the Heaviside step function.  We study $50$-$50$ binary mixtures of
particles with diameter ratio $1.4$ to inhibit crystallization
\cite{speedy,ohern_short,ohern_long,xu}.  The particles have equal
mass $m$ and are confined to a square unit cell with periodic boundary
conditions.  We chose the small particle diameter $\sigma$ and
$\sigma \sqrt{m/\epsilon}$ as the reference length and time scales.

Our focus is on configurations that are at a potential energy minimum
with infinitesimal overlaps.  Since these configurations are in
mechanical equilibrium and possess vanishingly small overlaps, we term
them mechanically stable (MS) packings.  MS packings with no overlaps
are equivalent to collectively jammed states \cite{torquato} of hard
disks.

We employ a class of packing-generation protocols that involve
successive compression or decompression steps followed by energy
minimization \cite{makse,xu}.  The process is initiated by choosing
random initial positions for the particles at packing fraction $\phi =
0.60$, which is well below the minimum packing fraction at which MS
packings occur in 2d.  The system is decompressed when the potential
energy of the system at a local minimum is nonzero; otherwise, the
system is compressed.  The increment by which the particle packing
fraction is changed at each compression or decompression step is
gradually decreased.  After a sufficiently large number of steps a MS
packing with infinitesimal overlaps is obtained.  This process is
performed for a large number of independent starting conditions to
generate an ensemble of MS packings.  In this way, we can measure the
probability to obtain a MS packing at a given $\phi$.

We employ two energy-minimization methods: (a) conjugate-gradient (CG)
minimization algorithm or (b) molecular dynamics (MD) with dissipation
proportional to local velocity differences.  The conjugate-gradient
method is a numerical scheme that begins at a given point in
configuration space and moves the system to the nearest local
potential energy minimum without traversing any energy barriers
\cite{numrec}.  In contrast, molecular dynamics with finite damping is
not guaranteed to find the nearest local potential energy minimum
since kinetic energy is removed from the system at a finite rate.  The
system can thus surmount a sufficiently low energy barrier.  In the
molecular dynamics method, each particle $i$ obeys Newton's equations
of motion
\begin{equation} \label{newton1}
m {\vec{a}}_i = 
\sum_{j\not=i}
   \Theta(\sigma_{ij}/r_{ij}-1) \left[\frac{\epsilon}{\sigma_{ij}}
\left(1-\frac{r_{ij}}{\sigma_{ij}} \right) - b\vec{v}_{ij} \cdot
{\hat{r}}_{ij} \right] {\hat{r}}_{ij},
\end{equation}
where $\vec{a}_i$ is the acceleration of particle $i$, $\vec{v}_{ij}$
is the relative velocity of particles $i$ and $j$, ${\hat{r}}_{ij}$ is
the unit vector connecting the centers of these particles, and $b = 0.5$ is
the damping coefficient.  Note that in our previous studies
\cite{ohern_short,ohern_long,xu} we used only the CG method.  

\subsection{Classification of Mechanically Stable Packings}
\label{classify}

In our numerical simulations we distinguish distinct mechanically
stable disk packings by the lists of eigenvalues of their dynamical
matrices.  For a pairwise additive, rotationally invariant potential
(\ref{spring potential}) the dynamical matrix is given by the
expressions \cite{tanguy}
\begin{equation}
\label{offdiagonal}
M_{i\alpha,j\beta}
  =-\frac{t_{ij}}{r_{ij}}(\delta_{\alpha \beta}
    -{\hat r}_{ij\alpha}{\hat r}_{ij\beta})
    -c_{ij} {\hat r}_{ij\alpha} {\hat r}_{ij\beta},
                 \quad i\not=j,
\end{equation}
and
\begin{equation}
\label{diagonal}
M_{i\alpha,i\beta} = - \sum_{j\not=i} M_{i\alpha,j\beta},
\end{equation}
where $t_{ij} = \partial V/\partial r_{ij}$ and $c_{ij} = \partial^2 V/
\partial r_{ij}^2$.  In the above relations the indices $i$ and $j$ refer to
the particles, and $\alpha,\beta=x,y$ represent the Cartesian coordinates.
For a system with $N_f$ rattlers and $N'=N-N_f$ particles forming a
connected network the indices $i$ and $j$ range from $1$ to $N'$, because the
rattlers do not contribute to the potential energy.  

The dynamical matrix is symmetric and has $dN'$ rows and columns,
where $d=2$ is the spatial dimension.  Thus it has $dN'$ real
eigenvalues $\{ m_i \}$, $d$ of which are zero due to translational
invariance of the system.  In a MS disk packing, no set of particle
displacements is possible without creating an overlapping
configuration; therefore the dynamical matrix has exactly $d(N'-1)$
nonzero eigenvalues.  In our simulations we use the criterion $|m_i| >
m_{\rm min} = 10^{-6}$ for nonzero eigenvalues.

We consider two MS packings to be the same if they have the same list
of eigenvalues of the dynamical matrix.  The eigenvalues are
considered to be equal if they differ by less than the noise threshold
$m_{\rm min}$ for our calculations.  Using the CG and MD methods, we
have identified $\approx 1600$ distinct MS packings for systems with $10$
particles.  (Packings with rattlers have been included in this count.)

It is in general not true that each distinct MS packing possesses a
unique packing fraction $\phi$.  However, we find that for these
systems only at most a few percent of distinct MS packings share the
same packing fraction.  Thus, in the following we will associate a
unique $\phi$ with each MS packing to simplify the discussion.

\begin{figure}
\centerline{\scalebox{0.5}{\includegraphics{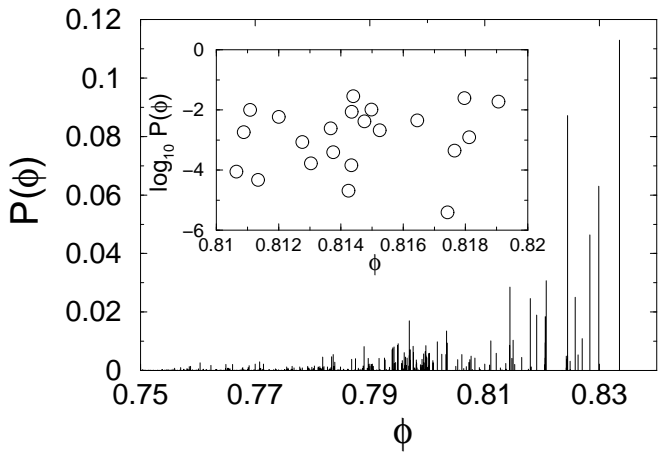}}}%
\vspace{-0.1in} 
\caption{\label{fig:TD} The probability to obtain a MS packing at
packing fraction $\phi$ for $N=10$ particles using the MD method. The
inset shows a magnified view of the probability on a logarithmic scale
over a narrow packing-fraction interval between $0.81$ and $0.82$.}
\end{figure}

\section{Results}
\label{results}

The principal result of this work is that the frequency with which MS
packings occur can vary dramatically---by many orders of magnitude
from one packing to another.  In Fig.~\ref{fig:TD}, we show the
probability $P(\phi) = n(\phi)/N_t$ to obtain a given MS packing at
$\phi$ for the MD energy minimization method.  $n(\phi)$ is the number
of MS packings obtained at $\phi$ out of $N_t$ trials performed.  For
$N=10$, we ran more than $30 \times 10^{6}$ trials for both energy
minimization methods and obtained results that do not depend on $N_t$.

We notice several interesting features in Fig.~\ref{fig:TD}.  First,
the probability distribution is not uniform.  The most probable MS
packings occur at large $\phi \approx 0.83$ and the least probable
occur for $\phi < 0.75$.  When averaged over narrow packing-fraction
intervals, probability distribution decays exponentially with
decreasing $\phi$ \cite{xu}.  Even more striking, the probability is
not monotonic in packing fraction and is in fact noisy and difficult
to predict.  In the inset to Fig.~\ref{fig:TD}, we find that even over
a narrow range $\Delta \phi$, the probability varies by more than five
orders of magnitude, and this occurs over the entire range of $\phi$.

\begin{figure}
\centerline{\scalebox{0.45}{\includegraphics{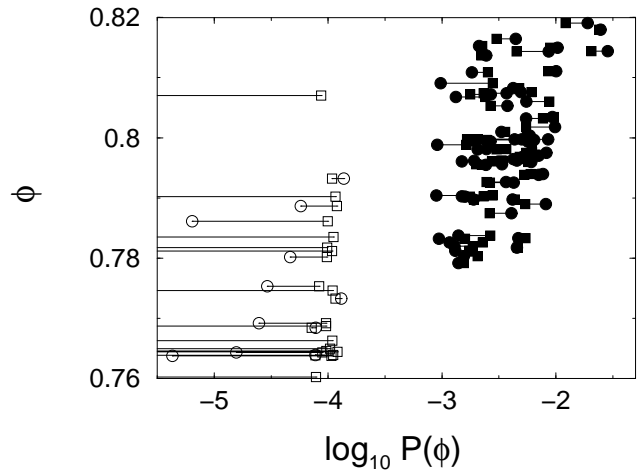}}}%
\vspace{-0.1in}
\caption{\label{fig:frequent} The probabilities for the most frequent 
MS packings obtained from the CG packing-generation algorithm (filled 
squares) are compared to the probabilities for the same MS packings
obtained from the MD method (filled circles).  A similar comparison
of less frequent MS packings obtained from the CG (open squares)
and MD (open circles) methods are also shown.}
\end{figure}

To understand the influence of the packing-generation protocol on our
results, we have examined the probabilities for obtaining each MS
packing using two energy minimization methods. We compared the $100$
most frequent MS packings obtained from the CG method to the $100$
most frequent packings from the MD method.  We find that $\sim 80\%$
of the MS packings were common to both sets; these are shown as filled
symbols in Fig.~\ref{fig:frequent}.  A similar comparison of
probabilities for obtaining less frequent MS packings using the MD and
CG methods is also displayed (using open symbols) in
Fig.~\ref{fig:frequent}.  We make several important observations.
First, significant shuffling of frequent and rare packings does {\it
not} occur---frequent packings remain frequent and rare ones remain
rare.  Second, within each probability grouping (i.e. filled versus
open symbols) the more frequent MS packings in the CG method become
more frequent in the MD method, while the less frequent packings from
the CG method become more rare in the MD method.  Although these
results were obtained for small systems, we expect similar findings
for large systems.

The result that frequent packings become more frequent and rare ones
become more rare when switching from CG to MD energy minimization can
be explained in part by considering the rate at which energy is
dissipated in the system. The MD energy minimization method utilizes a
finite rate of energy dissipation. Thus, at various points during the
packing generation process, the system can in principle possess enough
kinetic energy to jump out of a shallow basin corresponding to a rare
MS packing and into the basin of a more frequent packing. This
suggests that understanding the topography of configuration space
surrounding MS packings is crucial to understanding the frequency of
MS packings.

Also, the result that frequent MS packings remain frequent
and rare ones remain rare when switching from the MD to CG methods
implies that geometrical features of configuration space strongly
influence the frequency of MS packings.  Correlations between the
frequency with which MS packings occur and the shape and volume of
basins near each MS packing is a topic of our current investigations
\cite{gg}.

\section{Conclusion}
\label{conclusion}

We have enumerated nearly all of the distinct MS in small systems
composed of $50$-$50$ mixtures of bidisperse disks.  This has allowed
us to accurately measure the probability to obtain each MS packing.
We find that the probability is not uniform---packings with large
$\phi$ are exponentially more likely than those at low $\phi$.
Moreover, even over narrow packing-fraction intervals, distinct MS
packings occur with frequencies that differ by many orders of
magnitude.  We have varied the packing-generation protocol and shown
that these results do not change qualitatively, which suggests that
geometrical features of configuration space strongly influence the
packing frequencies. These results suggest that further work should be
carried out to explicitly test the Edwards' assumption that stable
grain packings are equally probable and similar assumptions about
inherent structures.  Another intriguing possibility is that only the
most frequent MS packings are relevant for slow dynamics in glassy and
jammed systems.  However, even if this were true, we must first
understand what determines MS packing frequencies in order to identify
correctly the relevant set of frequent MS packings.

\vspace{-0.2in}

\section*{Acknowledgements}

Financial support from NSF grant numbers CTS-0348175 (GG,JB),
DMR-0448838 (GG,CSO), and CTS-0456703 (CSO) is gratefully
acknowledged.  We also thank Yale's High Performance Computing Center
for generous amounts of computer time.


\begin{references}

\bibitem{stillinger}
F. H. Stillinger, {\it Science} {\bf 267}, 1935 (1995).

\bibitem{shell}
M. S. Shell and P. G. Debenedetti, {\it Phys. Rev. E} {\bf 69}, 051102 (2004).

\bibitem{nave}
E. LaNave, S. Mossa, and F. Sciortino, {\it Phys. Rev. Lett.} {\bf 88}, 
225701 (2002).
 
\bibitem{shell2}
M. S. Shell and P. G. Debenedetti, {\it J. Phys. Chem. B} {\bf 108}, 6772 
(2004).

\bibitem{edwards}
S. F. Edwards and R. B. S. Oakeshott, {\it Physica A} {\bf 157}, 1080 (1989).

\bibitem{makse2}
H. A. Makse and J. Kurchan, {\it Nature} {\bf 415}, 614 (2002).

\bibitem{xu}
N. Xu, J. Blawzdziewicz, and C. S. O'Hern, {\it Phys. Rev. E} {\bf 71}, 
061306 (2005). 

\bibitem{bs}
R. K. Bowles and R. J. Speedy, {\it Physica A} {\bf 262}, 76 (1999).

\bibitem{bsv}
R. K. Bowles and I. Saika-Voivod, {\it Phys. Rev. E} {\bf 73}, 011503 (2006).

\bibitem{adam}
G. Adam and J. H. Gibbs, {\it J. Chem. Phys.} {\bf 43}, 139 (1965).

\bibitem{gg}
G.-J. Gao, J. Blawzdziewicz, and C. S. O'Hern (unpublished).

\bibitem{torquato2}
S. Torquato, T. M. Truskett, and P. G. Debenedetti, {\it Phys. Rev. Lett.}
{\bf 84}, 2064 (2000).

\bibitem{speedy}
R. J. Speedy, {\it J. Chem. Phys.} {\bf 110}, 4559 (1999).

\bibitem{ohern_short}
C. S. O'Hern, S. A. Langer, A. J. Liu, and S. R. Nagel, {\it Phys. Rev. Lett.}
{\bf 88}, 075507 (2002).

\bibitem{ohern_long}
C. S. O'Hern, L. E. Silbert, A. J. Liu, S. R. Nagel, {\it Phys. Rev. E}
{\bf 68}, 011306 (2003).

\bibitem{torquato}
S. Torquato and F. H. Stillinger, {\it J. Phys. Chem. B} {\bf 105}, 11849 
(2001).

\bibitem{makse} H. A. Makse, D. L. Johnson, and L. M. Schwartz, {\it
Phys. Rev. Lett.} {\bf 84}, 4160 (2000); H. P. Zhang and H. A. Makse,
{\it Phys. Rev. E} {\bf 72}, 011301 (2005).

\bibitem{numrec} 
W. H. Press, B. P. Flannery, S. A. Teukolsky, and W. T. Vetterling, 
{\it Numerical Recipes in Fortran 77} (Cambridge University Press, 
New York, 1986). 

\bibitem{tanguy}
A. Tanguy, J. P. Wittmer, F. Leonforte, and J.-L. Barrat, 
{\it Phys. Rev. B} {\bf 66}, 174205 (2002).


\end{references}
\end{document}